\documentclass{article}

\usepackage{graphicx}
\usepackage[T1]{fontenc}
\usepackage{units}
\usepackage{mathcomp}
\usepackage{tikz}
\usetikzlibrary {decorations.pathmorphing}
\usepackage{url}
\usepackage{hyperref}

\title{Differential Temperature Anemometer}
\author{H. Nobach\\[1.5mm]{}\small Max Planck Institute for Dynamics and Self-Organization\\{}\small 
Am Fa\ss{}berg 17, 37077 G\"ottingen, Germany\\[1.5mm]{}\small holger.nobach@nambis.de}%
\date{\today}

\begin{document}

\maketitle

\begin{abstract}
Common thermal anemometers (hot-wire, hot-film, or similar) are based on the thermal equilibrium between the electrical power 
heating the sensor and the convection of the ambient medium cooling the sensor.
The response times of such instruments are often insufficient in rapidly fluctuating flow processes due to the thermal inertia 
of their sensing elements.
By using the instantaneous imbalance between the heating and the cooling power, which leads to a temperature gradient of the sensor, 
an instantaneous response of the measurement system can be achieved.
\end{abstract}

%
%

\section{Introduction}

In thermal anemometry a sensor (hot-wire, hot-film, or similar) is heated by electric current and cooled
by convection of a surrounding medium.
In thermal equilibrium and neglecting other heat losses, 
the measurement of the electrical heating power directly corresponds to the rate of convective heat flow.
With known caloric parameters of the ambient medium, this also yields information on the flow velocity.
Besides the density of the surrounding medium, its specific heat capacity and temperature difference to the 
heated sensor, the rate of convective heat flow also depends on the effective boundary layer thickness, which in turn 
is also a function of the flow velocity, unfortunately a non-linear one, which needs a calibration between the 
measured heating power and the corresponding flow velocity.

Constant voltage anemometers (CVA) and constant current anemometers (CCA) keep one electrical quantity 
contributing to the electrical power constant, either the voltage over the sensor element or the electrical current through it, 
while measuring the other quantity.
After a change of the flow velocity, a new equilibrium is reached at a different temperature.
Since temperature changes of the sensor element are slow, such anemometers lack temporal resolution.
In contrast, constant temperature anemometers (CTA) keep the sensor's temperature constant in an electrical control loop 
by providing exactly that electrical power to the sensor, which equals the rate of convective heat flow (plus possible additional losses).
Without changes of the sensor's temperature, the response time of the measurement corresponds to the response time of the driving control loop, 
trying to keep the sensor's temperature constant.
Since thermal anemometry is a well-established measurement method with a long history and many improvements over the years,
the literature is very extensive.
Reviews \cite{fingerson_94,thermalanemometry_handbook} give information on the various developments.
Appropriate investigations of the response functions of the various configurations of the instruments are available also \cite{bestion_etal_83}.

Instead of waiting for the thermal equilibrium after a change of the flow velocity, 
thermal transient anemometers (TTA) \cite{foss_etal_04} observe the temperature decay over time.
By fitting an exponential function, which is the stationary solution of the underlying differential equation, 
and extracting its time constant, the latter one also corresponds to the flow velocity after appropriate calibration.
With this, TTA can also be faster than conventional CVA or CCA, however, measurements with this method stop heating during 
the measurement, finally preventing this method from continuous operation.

In the present paper, a method is introduced, which also determines the temperature gradient of the sensor element,
however, here the underlying differential equation is used in its raw form.
By determining the instantaneous temperature gradient,
differential temperature anemometry (DTA) uses the instantaneous imbalance between electrical heating and convective cooling, 
finally yielding the intended information on the flow velocity with high temporal resolution.
The method can be based 
on existing thermal anemometer hardware like CVA, CCA or CTA, however, its principle also allows simpler hardware
without the need to keep any quantity constant during the measurement process.
In contrast to TTA, the sensor
can be heated continuously, allowing continuous velocity measurements like with conventional thermal anemometers.
A patent application has been filed through the Max Planck Society for the DTA method \cite{hwa_patentantrag}.

\section{Thermal anemometer technology and physical quantities}

\begin{figure}
\centerline{\includegraphics{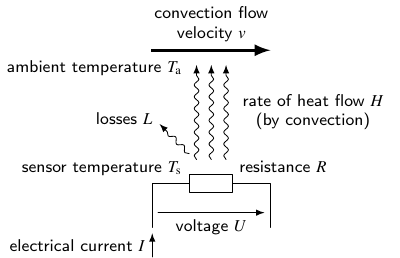}}
\caption{Thermal anemometry principle for measuring the flow velocity.\label{fig:hwasketch}}
\end{figure}

In thermal anemometry, a sensor element is heated by an electrical current $I$ through the sensor 
increasing its temperature $T_{\mathrm{s}}$ above the ambient temperature $T_{\mathrm{a}}$ (see Fig.~\ref{fig:hwasketch}).
The heating power is equal to the driving electrical power $P$, which is the product of the driving current $I$ 
and the voltage $U$ over the sensor.
From $U$ and $I$, also the resistance $R$ of the sensor can be obtained, yielding information about the sensor's temperature.
Depending on the material of the sensor element, different relations between $T_{\mathrm{s}}$ and $R$ apply.
The correspondence between $T_{\mathrm{s}}$ and $R$ can be derived either from the {\em a priori} known material properties 
or from an appropriate calibration.

On the other side, the sensor element is cooled by convection of the ambient medium at its temperature $T_{\mathrm{a}}$ 
and its flow velocity $v$.
The electrical heating power $P$ competes against the convective dissipation at the rate of heat flow $H$ away from the sensor.
In thermal equilibrium and neglecting other heat losses, $P$ and $H$ are balanced and the temperature of the sensor stays unchanged.
Changing the flow velocity, initially leads to an imbalance between $P$ and $H$ and to a temperature drift of the sensor.
The anemometer hardware can adjust the electrical power to keep the balance between $P$ and $H$ and subsequently 
also the temperature of the sensor.
Otherwise, the imbalance between $P$ and $H$ and the subsequent change
of the sensor's temperature will asymptotically achieve a new equilibrium at a different temperature of the sensor.

CVA and CCA keep one of the electrical quantities $U$ resp.\ $I$ constant and determine the new thermal equilibrium
by measuring the other electrical quantity.
Combining the two electrical quantities yields the electrical heating power in balance with the new rate of heat flow 
at the new equilibrium temperature of the sensor.
The change of the sensor's temperature due to the change of the flow velocity makes these anemometers slow due to an inefficient 
heat transfer from the sensor to the ambient medium.

CTA instead determines the temperature of the sensor and adjusts the electrical heating power into the sensor in a control loop
in that way that the sensor's temperature will stay constant.
Without slow changes of the sensor's temperature, CTA has the potential
to be much faster than the previous anemometers.
However, CTA still utilizes the equilibrium between $P$ and $H$ and the sensor
element is part of the temperature control loop.
A change of the flow velocity, therefore, will initially change the sensor's temperature, at least a bit.
Only then the control loop can invert the temperature deviation by adapting the heating power.
Therefore, the response time of CTA essentially depends on the quality and the speed of the control loop, 
which requires special effort and experience to be successful.

TTA instead determines temporal changes of the sensor's temperature during an imbalance between $P$ and $H$.
The underlying differential equation for the 
temporal gradient of the sensor's temperature $\dot T_{\mathrm{s}}=\frac{\mathrm{d}\,T_{\mathrm{s}}}{\mathrm{d}\,t}$
and the imbalance between $P$ and $H$ is
\begin{equation}
C_{\mathrm{s}}\dot T_{\mathrm{s}}=P-H\label{eq:dgl}
\end{equation}
with the heat capacity (thermal mass) of the sensor $C_{\mathrm{s}}$. 
By fitting an exponential function as its stationary solution and determining the time constant, 
TTA infers the flow velocity using appropriate calibration.
For this, the varying temperature of the sensor element is considered, which influences the relation between 
the rate of convective heat flow $H$ and the flow velocity $v$.
However, the use of the stationary solution of the differential equation implies constant boundary conditions during
the measurement process.
This includes both, the flow velocity and the heating.
For simplicity, the heating is stopped during the actual measurement process.
Therefore, TTA instruments are not operated continuously.

\section{New technique for fast response}

Instead of using the stationary solution of the underlying differential equation (\ref{eq:dgl}) of the heat balance, 
the new differential temperature anemometer (DTA) determines all contributing quantities of
the differential equation. 
To infer on an instantaneous rate of convective heat flow $H$
requires to acquire both, $P$ and $T_{\mathrm{s}}$ together with the intended temporal resolution. 
The instantaneous electrical heating power $P$ can be obtained directly from the primary electrical quantities $U$ and $I$ via
\begin{equation}
P=UI.
\end{equation}
For $T_{\mathrm{s}}$ first the electrical resistance $R$ of the sensor is determined as
\begin{equation}
R=\frac{U}{I}
\end{equation}
and translated into the corresponding temperature $T_{\mathrm{s}}(R)$ of the sensor based either on sufficient knowledge about
the specific resistance of the sensor's material, e.g.\ metals or semiconductors, or based on a dedicated temperature calibration
of the sensor's resistance.

However, the first temporal derivative $\dot T_{\mathrm{s}}$ of the sensor's temperature is also needed, which 
can be obtained from consecutive samples by numerical differentiation as
\begin{equation}
\dot T_{\mathrm{s}}=\frac{\Delta T_{\mathrm{s}}}{\Delta t}
\end{equation}
with the change of the sensor's temperature $\Delta T_{\mathrm{s}}$ observed over the sampling interval $\Delta t$.

The difference between the electrical heating $P$ and the heat accumulation of the sensor $C_{\mathrm{s}}\dot T_{\mathrm{s}}$ then 
yields the thermal loss by convection as
\begin{equation}
H=P-C_{\mathrm{s}}\dot T_{\mathrm{s}}-L(T_{\mathrm{s}}).\label{eq:baleq}
\end{equation}
The temperature dependent losses $L(T_{\mathrm{s}})$ have been found to influence the measurement significantly and have been added to the 
governing differential equation.
The observations in the present proof-of-concept experiment discovered a linear correspondence over the relevant temperature range.
Neglecting temperature differences within the sensor element and assuming the caloric parameters of the ambient medium being constant, 
the rate of convective heat flow $H$ itself linearly depends on the temperature difference between the sensor and the ambient medium.
\begin{equation}
H=(T_{\mathrm{s}}-T_{\mathrm{a}})g(v)\label{eq:veleq}
\end{equation}
In the first approximation, the specific calibration function $g(v)$ itself is independent of the actual temperature of the sensor 
and subsequently allows the procedure to work at drifting temperatures of the sensor.
For this, of course, the temperature of the ambient medium $T_{\mathrm{a}}$ is needed. 
It can either be kept constant or it is acquired together with the other quantities during the measurement.
However, large variations of the ambient temperature also lead to changes of the density and viscosity, 
which finally lead to deviations from the calibration function.
Note that this dependence of the rate of heat flow on the sensor's temperature and the calibration function $g(v)$ 
has also been used for TTA previously, while an explicit mathematical formulation is given elsewhere \cite{thermalanemometry_handbook}.


\section{Calibration}

The two equations, (\ref{eq:baleq}) and (\ref{eq:veleq}), for determining the intended flow velocity define the required
calibration procedures before a measurement.
\begin{enumerate}
\item The electrical heating power is determined from measuring the voltage $U$ over the sensor and the electrical current $I$ 
through the sensor.
Usual digital acquisition systems are optimized for accurate voltage measurements with a sufficiently high input resistance.
The electrical current through the sensor can be obtained e.g.\ via a voltage measurement 
over a shunt resistor in series with the thermal sensor.
An accurate measurement of the current then requires a
shunt resistor with a low temperature coefficient and the electrical resistance must be accurately known or measured.
The same calibration for the measurement of $U$ and $I$ is required for the determination of the sensor's resistance.

\item The determination of the heat imbalance between the sensor and the ambient medium needs both, the sensor's temperature
and its gradient.
The temperature can be obtained from the measurement of the sensor's resistance.
The relation depends on the material of the conductor and its geometry including the electrical contacts. 
Therefore, a calibration of the temperature dependent resistance of the particular sensor element is required.
However, this calibration is needed only once for the particular sensor and can be part of the manufacturing process.
The same holds for the measurement of the ambient temperature if done with another resistive probe.

\item Temperature dependent losses $L(T_{\mathrm{s}})$ have been found to have a significant influence on the measurement.
These losses are specific for the experimental setup.
Therefore, an appropriate calibration must be done within the experimental 
environment.
For that the temperature of the sensor can be measured at varying electrical heating power, while 
the other terms in Eq.~(\ref{eq:baleq}) are kept zero.
This is no convective heat flow, which is the case in 
still environment without a flow, and no temperature gradient, which means the sensor must reach its thermal equilibrium 
with the ambient for this calibration. 

\item The heat capacity of the sensor $C_{\mathrm{s}}$ can be obtained by observing the temperature gradient of the sensor $\dot T_{\mathrm{s}}$ 
during the transition after applying a certain heating power $P$ without a flow, such that the rate of heat flow $H$ into the ambient medium 
is zero and, only the temperature-dependent losses apply, which are known from the previous calibration step.
In this case 
\begin{equation}
C_{\mathrm{s}}=\frac{P-L(T_{\mathrm{s}})}{\dot T_{\mathrm{s}}}
\end{equation}
yields the heat capacity of the sensor.
This calibration can be repeated with modified heating power, yielding more robust estimates of the sensor's heat capacity.
This calibration is needed only once for the particular sensor and can be part of the manufacturing process,
unless the sensor gets contaminated with thermally affecting substances at later times.

\item Finally, the rate of heat flow $H$ is determined for various velocities $v$ by measuring the heating power $P$, 
while the wire temperature is constant, which means the sensor must reach its thermal equilibrium.
Fortunately, this is the same calibration procedure as for conventional thermal anemometer systems and, 
it can be performed with the same calibration devices,
where various defined flow velocities $v$ are provided and the appropriate system's output quantity is observed.
Without a temperature gradient, the above equations (\ref{eq:baleq}) and (\ref{eq:veleq}) yield the correspondence 
between the flow velocity $v$ and the reaction of the DTA processing $g(v)$ through
\begin{equation}
g(v)=\frac{P-L(T_{\mathrm{s}})}{T_{\mathrm{s}}-T_{\mathrm{a}}},
\end{equation}
where the previously obtained calibrations are used and, additionally the ambient temperature $T_{\mathrm{a}}$ is needed, 
like during the actual measurement.
\end{enumerate}

\section{Experimental proof of concept}

An experimental proof of concept should show that the proposed method is potentially faster than the commercial devices.
Therefore, a comparably slow sensor is used here instead of a fast hot-wire probe, namely a $\unit[2.3]{mm}\times\unit[2]{mm}$
large PT100 chip thin film resistor of the type S8P038 by Telemeter.
The measurement resistor is heated electrically and cooled in an air flow provided by a $\unit[12]{V}$ axial power supply fan 
of the type HEC FD128020HS with a diameter of $\unit[80]{mm}$, a rated speed of $\unit[2900]{rpm}$ and 
a rated flow rate of $\unit[23]{cfm}$.
The flow has been provided repeatedly to the PT100 resistor 
for $\unit[20]{s}$ and then blocked for another $\unit[20]{s}$ by putting a cup over the sensor. 
During the interval with blocked air flow, the fan speed has been increased and, then held constant for the next 
interval with air flow provided.

\begin{figure}
\centerline{\includegraphics{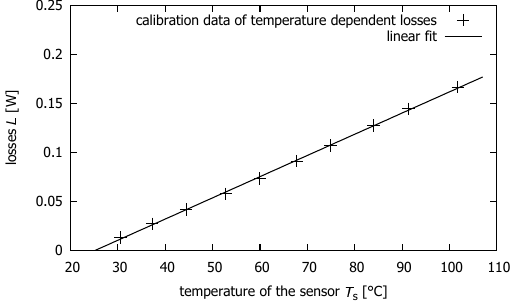}}
\caption{Calibration data and linear fit for temperature dependent losses.\label{fig:losses}}
\end{figure}

For deriving the sensor's temperature from its resistance, the known temperature coefficients of platinum
($\alpha=\unit[3.9083\times 10^{-3}]{K^{-1}}$ and $\beta=\unit[-5.775\times 10^{-7}]{K^{-2}}$)
have been used, without further calibration of the particular sensor.
The calibration data of temperature dependent losses shown in Fig.~\ref{fig:losses} yield a linear relation with a slope
of $\unit[2.16]{\frac{mW}{K}}$.
The heat capacity of the sensor has been found from the exponential temperature loss during time intervals without 
providing the flow to the sensor to be $C_{\mathrm{s}}=\unit[11.5]{\frac{mWs}{K}}$.

The thermal anemometer measurements have been done with three driving methods, a CCA, a CVA and a CTA, where
CCA is a separate circuit and CVA and CTA are realized with a combined circuit providing both options of driving.
During the measurements, the current through and the voltage over the thermal sensor have been recorded with a USB oscilloscope PicoScope 5443B.
For the current, the voltage has been recorded over a shunt resistor ($\unit[10.14]{\tcohm}$ for the CCA driver 
and $\unit[10.16]{\tcohm}$ for the CVA/CTA driver) in series with the thermal sensor.
For CCA and CVA, the acquisition of one varying electrical quantity would be enough to conclude on the heating power, 
since the other primary electrical quantity is either held constant or it can be uniquely derived from {\em a priori} measured values of the 
electrical circuit.
However, this proof-of-concept experiment was designed to demonstrate that the DTA method works for changing electrical driving conditions also. 
Therefore, both quantities $U$ and $I$ have been recorded independently in all three driving cases to show that 
the measurement concept does not require any electrical quantities kept constant.

\subsection{CCA circuit}

\begin{figure}
\centerline{\includegraphics{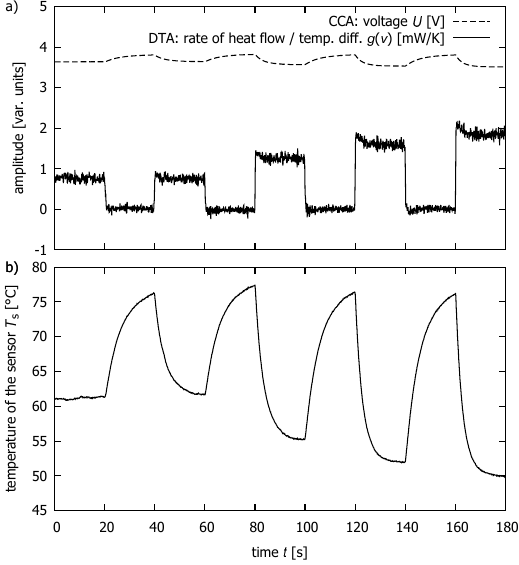}}
\caption{Data for proof of concept. a) CCA vs.\ DTA processing from the same experimental data 
using a CCA circuit and, b) sensor's temperature over time during the experiment.\label{fig:ccadata}}
\end{figure}

The CCA circuit provided a constant current of $I=\unit[29.4]{mA}$ through the thermal sensor.
The usual CCA measurement yields the voltage over the thermal sensor as its output quantity, which needs to be calibrated 
against the corresponding velocity of the ambient fluid.
For DTA the primary electrical quantities $U$ and $I$ have been processed further to the heating power and the sensor's resistance as 
$P=UI$ and $R=\frac{U}{I}$.
From the resistance of the sensor, its temperature $T_{\mathrm{s}}$ has been 
derived using an appropriate second-order polynomial correspondence for platinum with the two temperature coefficients given above.
Numerical differentiation of consecutive measurements of the sensor's temperature yields an estimate of the instantaneous temperature gradient
$\dot T_{\mathrm{s}}$. 

With these quantities, the instantaneous rate of heat flux $H$ has been obtained using Eq.~(\ref{eq:baleq}) and,
Eq.~(\ref{eq:veleq}) finally yields the DTA output quantity $g(v)=\frac{H}{T_{\mathrm{s}}-T_{\mathrm{a}}}$.
Fig.~\ref{fig:ccadata}a shows the respective output quantities for the measurement with the CCA circuit,
the voltage over the thermal sensor for usual CCA processing and the rate of heat flow per temperature difference 
for the new DTA processing.
Both curves have been derived from the same data recorded in one experiment.
Fig.~\ref{fig:ccadata}b additionally shows the temperature of the sensor during the experiment with varying flow velocities.
The appropriate calibrations to translate the quantities given by the two processing methods into 
flow velocities have not been done in this proof of concept, where response times were in focus.
Since the two measurement methods yield different quantities, the respective curves in Fig.~\ref{fig:ccadata}a do not correspond. 
Anyway, the strong low-pass damped character of the CCA processing is obvious, whereas the 
DTA processing provides a clear step-like response. 

The little overshoot in DTA, which increases with the step size of the flow provided is an indication
for the system to have two different time scales, where only the long-term time scale has been calibrated.
This also indicates that the first-order differential equation is not suited to describe the system comprehensively.
This makes sense, since the actual thermal thin-film sensor is supported by a thick ceramic substrate,
which introduces a big heat reservoir to the sensor. However, hot-wire probes don't have this and,
they should have more accurate responses than the sensor used in the present proof of concept.

\subsection{CVA circuit}

\begin{figure}
\centerline{\includegraphics{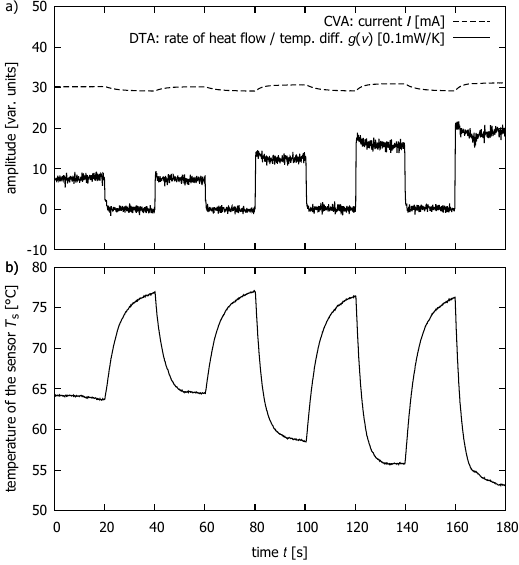}}
\caption{Data for proof of concept. a) CVA vs.\ DTA processing from the same experimental data 
using a CVA circuit and, b) sensor's temperature over time during the experiment.\label{fig:cvadata}}
\end{figure}

The combined CVA/CTA driver circuit provides a constant voltage, which then is supporting an inner controller loop
holding the sensor's temperature constant. 
By restricting the constant voltage and allowing arbitrary high temperatures, the voltage over the 
half bridge, consisting of the thermal sensor and the shunt resistor to obtain the current through the thermal sensor,
is kept constant at $U=\unit[4.07]{V}$. 
Keeping the voltage constant over the half bridge means that the voltage 
over the thermal sensor varies a bit depending on the current through the half bridge. 
This is not really a constant voltage anemometer and, the calibration curve is slightly different compared to the case where the 
voltage over the sensor is held constant. However, the shunt resistor is about one tenth of the sensor's resistance only, 
keeping the voltage fluctuations over the sensor moderate.
Furthermore, the temporal behavior of the CVA measurement remains unchanged anyway.
%
%
%
For DTA, again both electrical quantities, $U$ (namely the actual voltage over the sensor) and $I$ 
have been acquired and processed further to the heating power, the sensor's resistance, the sensor's temperature, 
its first temporal derivative, the rate of heat flux and finally to the DTA output quantity $g(v)$, which on the other hand 
provides information on the flow velocity, if calibrated appropriately. 

Fig.~\ref{fig:cvadata}a shows the respective output quantities for the measurement with the CVA circuit,
the current through the thermal sensor for usual CVA processing and the rate of heat flow per temperature difference 
for the new DTA processing.
Both curves have been derived from the same data recorded in one experiment.
Fig.~\ref{fig:cvadata}b additionally shows the temperature of the sensor during the experiment with varying flow velocities.
The appropriate calibrations to translate the quantities given by the two processing methods into 
flow velocities have not been done in this proof of concept, where response times were in focus.
Since the two measurement methods yield different quantities, the respective curves in Fig.~\ref{fig:cvadata}a do not correspond. 
Anyway, the strong low-pass damped character of the CVA processing is obvious, whereas the 
DTA processing provides a clear step-like response. 

Also here, the little overshoot in DTA after changing the flow velocity is obvious like with the previous CCA driver, 
indicating a higher-order damped characteristics with a heat reservoir in the back of the actual thermal sensor. 
The temporal characteristics are very similar between the two driving methods, CCA and CVA. 
For real hot-wire probes this phenomenon should not occur.

\subsection{CTA circuit}

\begin{figure}
\centerline{\includegraphics{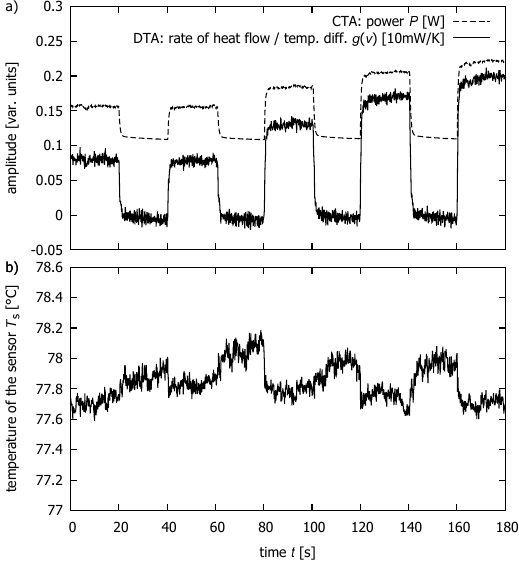}}
\caption{Data for proof of concept. a) CTA vs.\ DTA processing from the same experimental data 
using a CTA circuit and, b) sensor's temperature over time during the experiment.\label{fig:ctadata}}
\end{figure}

The CTA circuit uses a controller loop to keep the temperature of the sensor constant.
Fig.~\ref{fig:ctadata}b shows the remaining temperature fluctuations, which are insignificant compared to 
the previous two driving methods.
CTA yields the actual electrical heating power as its output quantity,
which is needed to keep the temperature constant. For that the primary electrical quantities measured, 
$U$ and $I$ are combined to the electrical power, which is shown in Fig.~\ref{fig:ctadata}a for the direct CTA
measurement.
Since slow temperature changes of the sensor are avoided, the response of the CTA is much faster 
than that of CCA or CVA.
However, appropriate calibration is needed to obtain the corresponding flow velocity
from the actual heating power.
Combining again the two primary electrical quantities further to
the sensor's resistance, the sensor's temperature, its first temporal derivative and the rate of heat flux, 
finally yields the DTA output quantity $g(v)$, which is also shown in Fig.~\ref{fig:ctadata}a.
The last one
can be translated into the actual flow velocity after appropriate calibration.
Since the appropriate calibrations for CTA and DTA to conclude on the flow velocity have not been done in this 
proof of concept, the respective curves in Fig.~\ref{fig:cvadata}a do not correspond. 
However, the temporal behavior can be compared.

The temporal behavior of the CTA circuit after a jump in the flow velocity deviates from the previous two driving methods, CCA and CVA.
Since CTA is not slowed down by temperature changes of the sensor, the response times of CTA and DTA compete.
Looking into details of the DTA response, the curves look like an underestimated heat capacity of the sensor on the first glance.
However, the insignificant temperature variations in Fig.~\ref{fig:ctadata}b show that this phenomenon cannot result from 
temperature changes of the resistive sensor directly.
On the other hand, the CTA curve in Fig.~\ref{fig:ctadata}a clearly shows a drift of the driving power during constant ambient flow conditions.
This means that e.g.\ the power 
to keep the temperature of the sensor element constant drifted downwards during the intervals without a flow.
On the other hand, 
this indicates that at the beginning of such an interval, an additional heat sink was effective, slowly losing its influence 
over time.
The ceramic substrate of the thin-film platinum resistor is a probable candidate leading to such behavior.
While the 
temperature of the thin-film resistor is held constant by the control loop, the temperature of the substrate may vary
during the experiment depending on the flow velocity.
Providing the flow to the entire sensor then leads to a temperature decrease
of the substrate, which at the beginning of the following interval without the flow is draining heat from the platinum sensor, 
and {\em vice versa}.
Combining the information
from the usual CTA processing and that of the DTA processing leads to the conclusion, that the drifts of the two 
output quantities observed fit together and are real losses of heat into a heat reservoir, which is not considered 
in the assumed governing differential equation (\ref{eq:baleq}).
In this case, real hot wires are predicted to 
have no such effect due to their construction, which has no such heat reservoir.

\section{Conclusion}

The results proof, that the proposed DTA method for thermal anemometers based on the instantaneous determination 
of all relevant quantities of the underlying differential equation has the potential to lead to faster responses than common thermal anemometers.
Since changes of the electrical heating during the measurement are registered during the measurement process, various driving 
concepts and circuits can be used, including common CCA, CVA or CTA circuits.
However, DTA does not require any driving quantity 
held constant.
Therefore, much simpler driving circuits can be used, yielding maximum flexibility in constructing the 
driving circuit.
The same holds for the thermal sensors, which may have various 
resistance ranges or non-linear responses and, the method works for both, thermal sensors with positive as well with negative 
temperature coefficients.

However, it should be mentioned, that the data sequence registered with the USB oscilloscope has a significant noise influence 
and needed massive oversampling and averaging for the processing in this paper.
The primary data have been obtained with $\unit[100]{kHz}$, 
whereas the data sequence, which has been used for the further processing has been reduced to $\unit[10]{Hz}$ sampling rate. 
Obviously, optimizing the circuit towards minimum noise introduction is a rewarding goal.
This and the investigation of real hot-wire probes without the disturbing heat reservoir of the present proof-of-concept experiment
will be the next steps for this promising development.
The data that support the findings of this study are openly available at \url{http://www.nambis.de/publications/arxiv23c.html}.

\bibliography{/Users/hnobach/Backup/Private/Tools/TEXINPUT/bibtex/bib/lite,/Users/hnobach/Backup/Private/Tools/TEXINPUT/bibtex/bib/myown,/Users/hnobach/Backup/Private/Tools/TEXINPUT/bibtex/bib/aktu}
\bibliographystyle{unsrt}

\end{document}